\documentclass[aps,prd,nofootinbib,twocolumn,groupedaddress]{revtex4}
\usepackage{graphicx}
\usepackage{amsmath,amssymb}
\usepackage{hyperref}
\usepackage[usenames]{color}
\usepackage{bm}

\hypersetup{
    colorlinks=true,
    linkcolor=red,
    citecolor=blue,
}

\def\be{\begin{equation}}
\def\ee{\end{equation}}
\def\ba{\begin{eqnarray}}
\def\ea{\end{eqnarray}}

\newcommand\araa{Annual Review of Astron and Astrophys}

\expandafter\ifx\csname natexlab\endcsname\relax\fi
\expandafter\ifx\csname bibnamefont\endcsname\relax
  \def\bibnamefont#1{#1}\fi
\expandafter\ifx\csname bibfnamefont\endcsname\relax
  \def\bibfnamefont#1{#1}\fi
\expandafter\ifx\csname citenamefont\endcsname\relax
  \def\citenamefont#1{#1}\fi
\expandafter\ifx\csname url\endcsname\relax
  \def\url#1{\texttt{#1}}\fi
\expandafter\ifx\csname urlprefix\endcsname\relax\fi
\providecommand{\bibinfo}[2]{#2}
\providecommand{\eprint}[2][]{\url{#2}}

\begin{document}

\title{Galactic Faraday rotation effect on polarization of 21cm lines from the epoch of reionization}

\author{Soma De$^{1}$ and Hiroyuki Tashiro$^{2}$ }

\affiliation{$^1$School of Earth and Space Exploration, Arizona State University, Tempe, AZ 85287, USA  \\
$^2$Physics Department, Arizona State University, Tempe, AZ 85287, USA  
}

\begin{abstract}
Redshifted 21~cm signal from neutral hydrogen is one of the most competitive probes
of the epoch of reionization (EoR). Unpolarized 21~cm
radiation acquires a certain level of linear polarization during the EoR 
due to Thompson scattering.
This linear polarization, if measured, could probe important information
about the EoR. 
We study the effect of Galactic Faraday rotation (FR) in the detection of polarization of
21~cm radiation. Since the effect of FR is strong due to the large
wavelength nature of redshifted 21~cm signal, the initial E-mode signal
carried by 21~cm photons is modified in its entirety. We show that a 99\% accuracy on rotation
measure (RM) data is necessary to recover the initial E-mode signal. We
conclude that, given the current status of RM observations,
it is not possible to retrieve the initial polarization signal of 21~cm
from the EoR. However, we are optimistic that it may be possible from the next
generation radio surveys. 
\end{abstract}
\maketitle

\section{Introduction}
 During the expansion history of our Universe, hydrogen went through
different phase transitions. Following nucleosynthesis in the early
universe, hydrogen and
helium were ionized, and then upon rapid cooling due to expansion,
hydrogen became neutral releasing the cosmic microwave background (CMB)
radiation.  This is known as the epoch of recombination which happened
around $z=1080$~\cite{hucmb2008}. Past this epoch, the Universe was neutral.
Later on high energy photons
released by the first stars and quasars again ionized the hydrogen in
the inter galactic medium (IGM). This phase is known as the epoch of
reionization (EoR).  Therefore the expansion history of the universe is
encoded in the phases of hydrogen, the most abundant element in the
Universe. Probing reionization is a way to
learn more about the first stars and first galaxies. The 21~cm line
corresponds to the ground state hyperfine transition of atomic hydrogen (see review by
Ref.~\cite{pritchard21cm2012}). The condition for detectability of such a
signal with respect to background CMB
depends on the excitation temperature of the 21 cm transition or
the spin temperature. If the spin temperature, $T_S$ deviates from the
CMB temperature, a signal may be detectable. This deviation from the
background CMB temperature is quantified in terms of the brightness
temperature~\cite{morales21cm2010} which directly depends on the
physical properties of IGM such as density, velocity gradients, gas
temperature, level populations and ionization stages.

The statistic of 21~cm brightness temperature fluctuations during the EoR
is described by the power spectrum in the Fourier space. This 21~cm
power spectrum directly depends on the power of dark matter density
fluctuations, baryonic ionization fluctuations and relevant cross
terms. Therefore power spectrum plays a crucial role in distinguishing
between several models of re ionization dynamics~\cite{Masui21cm2013,
Kovetz21cm2013}. 
There have been several papers
discussing how the brightness temperature and power spectra of 21~cm
could enlighten us on the EoR.

In this paper, we discuss the polarization of the cosmological 21~cm
signal and Faraday rotation (FR) which plays a major
foreground in its detection.
Although the major cosmological 21~cm signal is intrinsically unpolarized,
as discussed in Ref.~\cite{babich21cm2005},  polarization in 21~cm radiation
could arise from the properties of the sources (e.g.~due to Zeeman
effect \cite{cooray21cm2005}) and primarily Thomson scattering in the
EoR.
The first polarization sources are very small, although Thomson
scattering can create significant level of polarization. The power
spectrum of the polarization created through Thomson scattering depends
on the re ionization physics such as the number density and the typical
size of ionized bubbles. Therefore, the polarization contain
complementary information on the EoR to the brightness temperature power
spectrum. In particular, Ref.~\cite{babich21cm2005} pointed out that the
peak scale of polarization power spectrum corresponds to the typical
scale of ionized bubbles at the epoch when observed 21~cm radiation is emitted.
The detection prospect of any particular type of polarization
will depend on the presence of foregrounds. Since its a low-frequency
signal, Faraday rotation is a crucial foreground with a definite
frequency dependence.

There are several papers for the effect of Galactic Faraday rotation on
the measurement of 21~cm signal~\cite{Jelic:2010vg, Geil:2010ud, Moore:2013ip}.
The Faraday rotation causes the leak of polarized foreground into the
estimate of the intrinsically unpolarized signal. Therefore, Galactic
Faraday rotation has been studied by using numerical simulation with
Galactic Faraday rotation model, in order to estimate
a contamination effect of polarized foreground 
 in these papers.

In this paper, we discuss how Galactic FR transforms the initial linear polarization of 21~cm signal
originating around the EoR. A calculation on how the linear polarization
of CMB is modified as it passes through the Milky Way was done in
Ref.~\cite{Decmb2013}. For the case of CMB,
although some B-mode polarization is produced due to the Galactic FR
effect, the change in the initial E-mode signal is negligible. However,
the angle of rotation in the case of 21~cm radiation
is large due to the larger wavelength nature of 21~cm photons from EoR. To take into account of large FR, we build a framework in this paper to calculate the
exact polarization spectra under any rotational angle. We then estimate
the effect of Galactic Faraday rotation on the initial-E mode polarization coming from EoR.

  The organization of the paper is as follows.
In Sec.~\ref{sec:frbasics} we discuss the basic principles of Faraday
rotation and how it transforms the polarization tensor. In
Sec.~\ref{sec:emodepol},
we discuss briefly how the E-mode polarization
originated in 21 cm during the EoR. 
In Sec.~\ref{sec:results}, we present the Faraday rotation map of
our Galaxy and quantify the effect  Faraday rotation due to Milky way
upon 21 cm E-mode polarization. We conclude our paper by discussing if
at all 21 cm polarization may be detected in future experiments and
acknowledgments.
Throughout this paper, we use cosmological parameters for a flat
$\Lambda$CDM model: $h = 0.73$ ($H0 = h \times 100$ km/s/Mpc),
$\Omega_c h^2 =0.104 $ and $\Omega_bh^2=0.022$.

\section{Basics of Faraday Rotation}
\label{sec:frbasics}
If photons emitted at a given redshift, $z$, with a rest-frame
wavelength of 21 cm pass through ionized regions permeated by magnetic
fields, the direction of $linear$ 
polarization is rotated by an angle 
\begin{equation}
\alpha(\hat{\bm n}) = \lambda_0^2 \ RM(\hat{\bm n})= \frac{3}{{16 \pi^2 e}} \lambda_0^2 \int \dot{\tau} \ {\bm B} \cdot d{\bm l} \ ,
\label{alpha-FR}
\end{equation}
where $\hat{\bm n}$ is the direction along the line of sight, $\dot{\tau}$ is the differential optical depth, 
$\lambda_0$ is the observed wavelength of the 21~cm radiation at
redshift $z=0$, ${\bm B}$ is
the ``comoving'' magnetic field, and $d{\bm l}$ is the comoving length
element along the photon trajectory. The rotation measure, $RM$, is a
frequency independent quantity in Eq.~(\ref{alpha-FR}) that describes the
strength of Faraday rotation. The observed Stokes parameters upon Faraday rotation angle
$\alpha (\hat n)$ along a chosen direction $\hat n$ are given by
\begin{equation}
    \left( \begin{array}{c} Q_{\mathrm{obs}} (\hat n) \\
     U_{\mathrm{obs}} (\hat n) \\ \end{array}
     \right)
     = \left( \begin{array}{cc} \cos 2
     \alpha(\hat n) & \sin 2\alpha(\hat n) \\ - \sin 2\alpha(\hat n)
     & \cos 2 \alpha(\hat n) \\ \end{array} \right)
     \left( \begin{array}{c} Q_i(\hat n) \\ U_i (\hat n) \\ \end{array}
     \right),\label{eq:rotation_polarization} 
\end{equation}
where $(Q_i, U_i)$ is the initial Stokes parameters prior to rotation.

The polarization tensor is expressed in terms of the Stokes parameter as
\begin{equation}
P_{\alpha \beta} (\hat n) = \left( \begin{array}{cc}
     Q(\hat n) & U(\hat n)\\ U(\hat n)
     & - Q(\hat n) \\ \end{array} \right).\label{polar_tensor}
\end{equation}

The polarization tensor
is a symmetric trace-free $2 \times 2$ tensor field.
Therefore, 
the polarization tensor can be expanded in terms of tensor spherical
harmonics as \cite{Kamionkowski:2008fp}
\begin{equation}
     {\cal P}_{ab} 
      (\hat n) = \sum_{\ell=2}^\infty\sum_{m=-\ell}^\ell
     \left[ E_{\ell m} 
     Y_{(\ell m)ab}^{\rm E}(\hat n) + B
     _{\ell m} Y_{(\ell m)ab}^{\rm B}
     (\hat n) \right],
\label{Pexpansion}
\end{equation}
where $Y_{(\ell m)ab}^{\rm E}(\hat n) $ and $Y_{(\ell m)ab}^{\rm B}(\hat n) $
are complete sets of basis functions for the E-mode and B-mode
components of the polarization~\cite{Kamionkowski:1996zd, Zaldarriaga:1996xe}.
Orthonormality of the basis functions provides
E- and B-mode 
coefficients of polarization as
\begin{eqnarray}
     E_{\ell m}  &=&\int \, d\hat n {\cal P}_{ab} (\hat n)
                             Y_{(\ell m)}^{{\rm E} \,ab\, *}(\hat n),
                             \nonumber  \\
     B_{\ell m} &=&\int d\hat n\, {\cal P}_{ab} (\hat n)
                                      Y_{(\ell m)}^{{\rm B} \, ab\, *}(\hat n).
\label{eqn:defmoments}
\end{eqnarray}
%
%

Using Eqs.~(\ref{eq:rotation_polarization}), (\ref{polar_tensor}) and
(\ref{eqn:defmoments}),
we obtain
\begin{flalign}
E_{\ell m} ^{\rm obs} &= \sum_{\ell' m'} \sum _{LM} \int \, d\hat n ~
Y_{(\ell'm')}^{{\rm E} \,ab\, *}(\hat n) Y_{(LM)} (\hat n)
 \nonumber \\
& \times \left [C_{LM} \left(E_{\ell'm'} ^{i} Y_{(\ell'm')ab}^{\rm E}(\hat n)
+ B_{\ell 'm'}^{i} Y_{(\ell 'm')ab}^{\rm B}(\hat n) \right) \right.
\nonumber \\
&  \left. +S_{LM} \left(B_{\ell 'm'} ^{i} Y_{(\ell 'm')ab}^{\rm E}(\hat n) 
+ E_{\ell 'm'} ^{i}Y_{(\ell 'm')ab}^{\rm B}(\hat n)   \right)
 \right],
\nonumber \\
B_{\ell m} ^{\rm obs}  &=
\sum_{\ell' m'} \sum_{LM} \int \, d\hat n ~ Y_{(\ell 'm')}^{{\rm B} \,ab\, *}(\hat n) Y_{(LM)} (\hat n)
\nonumber  \\
& \times \left [ 
      C_{LM} \left(E_{\ell 'm'} ^{i} Y_{(\ell 'm')ab}^{\rm E}(\hat n)
+ B_{\ell 'm'} ^{i} Y_{(\ell 'm')ab}^{\rm B}(\hat n)\right) \right.
 \nonumber \\
&\left. + S_{LM} \left (B_{\ell 'm'} ^{i} Y_{(\ell 'm')ab}^{\rm E}(\hat n) +
 E_{\ell 'm'} ^{i}Y_{(\ell 'm')ab}^{\rm B}(\hat n)   \right)
\right ]
 \label{eq:elm_blm_fara}
\end{flalign}
where $C_{\ell m}$ and $S_{\ell m}$ is the spherical harmonic coefficients
of $\cos 2 \alpha(\hat n)$ and $\sin 2\alpha(\hat n)$,
\begin{eqnarray}
 \cos 2 \alpha(\hat n) &=& \sum_{\ell m} C_{\ell m}
Y_{(\ell m)}(\hat n),
\nonumber \\
\sin 2\alpha(\hat n) &=& \sum_{\ell m} S_{\ell m}
Y_{(\ell m)}(\hat n).
\end{eqnarray}

As we discuss later, 
we consider Thomson scattering in the EoR
as a source of the 21~cm polarization.
Since Thomson scattering creates the only E-mode, we assume that
our initial polarization has $E^i_{\ell m} \neq 0 $ and $B^i_{\ell m} = 0 $ hereafter.

The integrations in
Eq.~(\ref{eq:elm_blm_fara})
can be performed as~\cite{Hu:2000ee}
\begin{flalign}
&\int \, d\hat n ~
Y_{(\ell' m')ab}^{\rm E}(\hat n)   Y_{(\ell' m')}^{{\rm E} \,ab\, *}(\hat n)
Y_{(LM)} (\hat n) \nonumber \\
& = \int \, d\hat n ~
Y_{(\ell ' m')ab}^{\rm B}(\hat n)   Y_{(\ell ' m')}^{{\rm B} \,ab\, *}(\hat n)
Y_{(LM)} (\hat n) =H^{L}_{\ell \ell '} \xi^{LM}_{\ell m \ell^\prime m^\prime},
\nonumber\\
&\int \, d\hat n ~
Y_{(\ell ' m')ab}^{\rm B}(\hat n)   Y_{(\ell 'm')}^{{\rm E} \,ab\, *}(\hat n)
Y_{(LM)} (\hat n) \nonumber \\
& =\int \, d\hat n ~
Y_{(\ell 'm')ab}^{\rm E}(\hat n)   Y_{(\ell 'm')}^{{\rm  B} \,ab\, *}(\hat n)
Y_{(LM)} (\hat n) = 0,
\label{eq:integration_eqs}
\end{flalign}
where
\begin{equation}
      H^{L}_{\ell \ell'} \equiv \left( \begin{array}{ccc} \ell & \ell' & L
     \\ 2 & -2 & 0 \\ \end{array} \right)
     \left( \begin{array}{ccc} \ell & \ell' & L
     \\ 0 & 0 & 0 \\ \end{array} \right)^{-1},
\end{equation}
in terms of Wigner-3j symbols, and
\begin{equation} \label{E:xint}
     \xi^{LM}_{\ell m \ell^\prime m^\prime} =
     \int d\hat n \, Y_{(\ell m)}^\ast(\hat n)
     Y_{(\ell^\prime m^\prime)}(\hat n)
     Y_{(LM)}(\hat n).
\end{equation}
The integrations of Eq.~(\ref{eq:integration_eqs})
are nonzero only for $\ell + \ell '+L =$even. 


Now we can write the observed E-mode and B-mode after Faraday rotation as
\begin{flalign}
E_{\ell m}^{obs}&=\sum_{\ell ' m'}\sum_{LM}
C_{LM}  \xi^{LM}_{\ell m \ell ' m'}H^{L}_{\ell \ell '} E^i _{\ell' m'},
\nonumber \\
B_{\ell m}^{obs}&=\sum_{\ell 'm'}\sum_{LM}
 S_{LM}\xi^{LM}_{\ell m \ell 'm'}H^{L}_{\ell \ell'} E^i _{\ell' m'}.
\end{flalign}

Accordingly, observed E-mode
and B-mode autocorrelation can be expressed as 
\begin{flalign}
&<E^{obs}_{\ell 'm'}E^{obs}_{\ell m}>\delta_{\ell \ell'}\delta_{mm'}\nonumber \\
&=\sum_{L}\left(\frac{2L+1}{4\pi}\right) C_{L}^{\cos(2\alpha)}\sum_{\ell '}(2 \ell'+1)
 C_{\ell'} ^{E,i}
 (H^{L}_{\ell \ell'})^2,
 \label{eq:somae} \\
&<B^{obs}_{\ell 'm'}B^{obs}_{\ell m}>\delta_{\ell \ell '}\delta_{mm'} \nonumber\\
&=\sum_{L} \left(\frac{2L+1}{4\pi} \right)C_{L}^{\sin(2\alpha)}\sum_{\ell'}(2 \ell'+1)C_{\ell'}^{E,i}
 (H^{L}_{\ell \ell'})^2.
 \label{eq:somab} 
\end{flalign}
In Eqs.~(\ref{eq:somae}) and (\ref{eq:somab}), $C_{\ell}^{E,i}$ indicates the
initial E-mode angular power spectrum, and $C_{\ell}^{\sin(2\alpha)}$ and
$C_{\ell}^{\cos(2 \alpha)}$ are the angular power spectra of
$\sin(2\alpha)$ and $\cos(2\alpha)$, respectively.


\section{21~cm E-mode polarization from the EoR}
\label{sec:emodepol}
Similarly to CMB polarization, 21~cm polarization is generated in the EoR~\cite{babich21cm2005}.
The fluctuations of emitting neutral hydrogen produce quadrapole temperature
anisotropy of free streaming redshifted 21~cm radiation around free
electrons in the EoR.
Thomson scattering with this quadrapole anisotropy radiation generates a
linear polarization.
The generated 21~cm polarization can be
decomposed into E- and B-modes~\cite{Kamionkowski:1996zd, Zaldarriaga:1996xe}.
Since the fluctuations of neutral hydrogen are due to the baryon and ionization
fraction fluctuations, 
the generated polarization is only E-mode,
due to the parity invariance.

Following the total angular momentum representation of CMB polarization~\cite{Hu:1997hp},
the E-mode polarization power spectrum of the redshifted 21~cm
fluctuations can be expressed as 
\cite{babich21cm2005}
\begin{eqnarray}
C^{E,i}_\ell (\nu) & =& \frac{2}{\pi} \int k^2 dk ~
\left[\bar{x}^2_H P_{\delta}(k)~ 
 \left(_{\delta}\Delta^E_{\ell }(k,\nu)\right)^2
\right.
 \nonumber \\
 && \qquad \left.
+ P_{x}(k)~\left(
_{x}\Delta^E_{\ell}(k,\nu)\right)^2\right],
\label{eq:cl_E_EOR}
\end{eqnarray}
where $\bar{x}_H$ is the average ionization fraction,
$P_{\delta}(k)$ is the baryonic power spectrum and $P_x(k)$ is the
ionization fraction power spectrum.

In Eq.~(\ref{eq:cl_E_EOR}), ${_{\delta}\Delta^E_{\ell }(k,\nu)}$ and
${_{x}\Delta^E_{\ell}(k,\nu)}$ are the transfer functions of the
quadrapole component  for the baryon fluctuations and ionization fraction
fluctuations, respectively,
\begin{flalign}
_{\delta}\Delta^E_{\ell}(k,\nu) &=
   \frac{3}{4}\sqrt{\frac{(\ell+2)!}{(\ell-2)!}} \int^{\eta_R}_0 d\eta 
\nonumber \\
 & \quad \times  \frac{g(\eta)}{\eta^2k^2} j_{\ell}(k\eta)
 \left[~j_2(k(\eta_E-\eta)) - j_2''(k(\eta_E-\eta)) \right],  
\\
 _{x}\Delta^E_{\ell}(k,\nu) &= \frac{3}{4}\sqrt{\frac{(\ell+2)!}{(\ell-2)!}} \int^{\eta_R}_0 d\eta 
   \frac{g(\eta)}{\eta^2k^2} j_{\ell}(k\eta) j_2[k(\eta_E-\eta)],  
\end{flalign}
where $\eta_E$ is the conformal time at the emission redshift.

For the case of $z_E \sim z_r $, the polarization due to ionization fraction fluctuations 
dominates the one due to the baryon fluctuations~\cite{babich21cm2005}.
Therefore, we take into account only $P_x$ contributions hereafter.

The power spectrum $P_x$ depends on the reionization model. 
We calculate $P_x$ following the way of \cite{babich21cm2005},
in which, although they considered only Poisson fluctuations,
$P_x$ is evaluated robustly.

First, we model the average ionization fraction to be a function of redshift
as used in \texttt{CAMB}~\cite{Lewis:1999bs}
\begin{equation}
\bar x_e(z) = \frac{1}{2} \left [
           1+{\rm tanh}\left ( \frac{2 (1+z_r)^{3/2} -(1+z)^{3/2} }{3\Delta
			z  (1+z_r)^{1/2}} \right )
                      \right ],
\label{wmapmodel}
\end{equation}
with $\Delta z \sim 0.5$.

Around the end of reionization, the bubble starts to
overlap~\cite{Wyithe:2004qd}.
Therefore, the mean distance between ionized bubbles
 corresponds to the characteristic size of the bubbles.
We define the averaged number density of ionized bubbles at $z_r$ as
\begin{equation}
 \bar{n} = \frac{3}{4\pi} R_r^{-3} ,
\end{equation}
where $R_e$ is the characteristic size of the bubbles at $z_r$.
During the reionization process, the characteristic scale of ionized
bubbles starts to grow from a scale smaller than $R_r$, grows with time
and finally reach $R_r$.
To take into account the growth of the characteristic scale, 
we simply assume that the average bubble density is constant and
the characteristic scale of bubbles, $R(z)$, evolves to satisfy
\begin{equation}
 \bar x_e (z) = \frac{4 \pi}{3} \bar n R(z)^3.
\end{equation}

Assuming that ionized bubbles are point-like objects,
the two-point correlation function for Poisson contribution
is given by
\begin{equation}
  \langle \delta x_p(\bm{r}_1) \delta x_p(\bm{r}_2) \rangle =
\frac{1}{\bar{n}(\bm{r}_1)} \delta^{(3)}(\bm{r}_1-\bm{r}_2),
\end{equation}
where $\delta x_p$ 
is the fluctuations of the ionization fraction due to the point-like
objects and we assume that the inside of ionized bubbles are totally ionized.
However, actual ionized bubbles have a finite size $R$. In order to consider the finite
size effect, we take the convolution with a window function,
\begin{equation}
   \delta x(\bm{r}_1) = \int d^3\bm{r}_2 W_R(|\bm{r}_1 - \bm{r}_2|) \delta x_p(\bm{r}_2),
\end{equation}
where we adopt a Gaussian window function with the window size $R$,
\begin{equation}
   W_R(r) = \frac{e^{-r^2/2R^2}}{\sqrt{(2\pi)^3} R^3}.
\end{equation}

Therefore, we can write the two-point correlation function of the
ionized fraction as
\begin{eqnarray}
&& \langle \delta x(\bm{k}_1) \delta x^{*}(\bm{k}_2) \rangle = 
\int d^3 \bm{r}_1 d^3 \bm{r}_2 e^{i \bm{k}_1 \cdot \bm{r}_1} e^{-i
\bm{k}_2 \cdot \bm{r}_2}
\nonumber \\
 && \quad
\times
\int d^3 \bm{y}_1  d^3 \bm{y}_2 
\frac{e^{-|\bm{r}_1 - \bm{y}_1|^2/2R^2}}{\sqrt{(2\pi)^3} R^3}
\frac{e^{-|\bm{r}_2 - \bm{y}_2|^2/2R^2}}{\sqrt{(2\pi)^3} R^3}
\nonumber \\
 && \quad \times
\langle \delta x_p(\bm{y}_1) \delta x_p(\bm{y}_2) \rangle.
\end{eqnarray}

Defining the power spectrum of the ionization fraction $P_x(k)$ as
$\langle \delta x(\bm{k}_1) \delta x^{*}(\bm{k}_2) \rangle = (2\pi)^3 \delta^{(3)}
(\bm{k}_1 - \bm{k}_2) P_x(k_1)$,
we obtain the power spectrum,
\begin{equation}
P_x(k) = \frac{1}{\bar{n}} e^{-k^2 R^2}.
\end{equation}
Accordingly, the peak scale of the power spectrum corresponds to the
typical scale of ionized bubbles in our model.


\begin{figure}[tbp]
\includegraphics[height=0.65\textwidth]{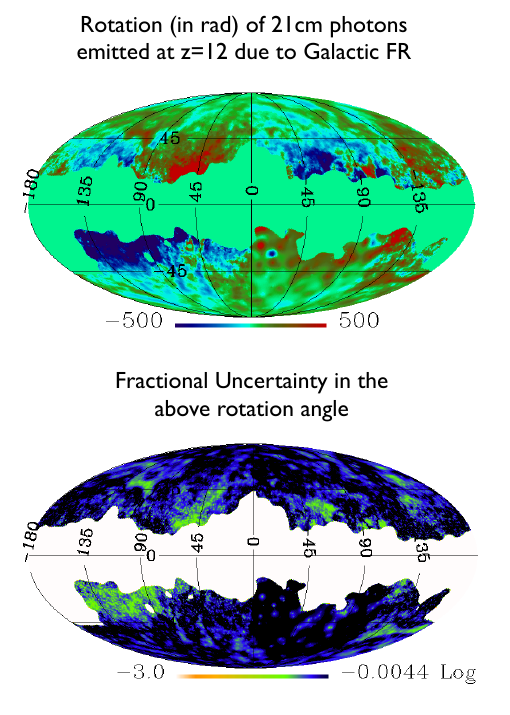}
\caption{
The Faraday rotation angle in radian of the rest frame 21~cm photons
from $z=$12 (top panel) and
the fractional uncertainty in the rotation angle~(bottom panel).}
\label{fig:anglemap}
\end{figure}

\begin{figure*}[tbp]
\includegraphics[height=1.0\textwidth]{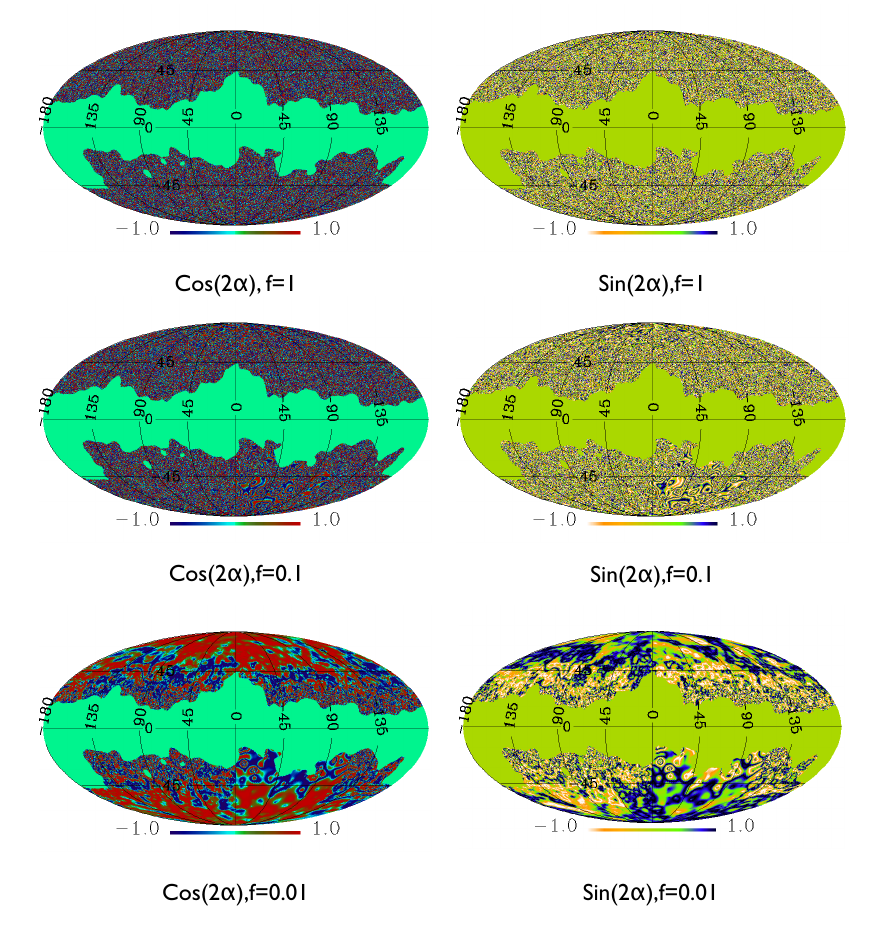}
\caption{Faraday rotation maps.
Top panel shows Cosine and Sine of twice the angle of rotation
as 21~cm photons emitted at $z=$12 pass through the galaxy. Mid and
bottom panels respectively indicate where RM is respectively
0.1 and 0.01 of the actual value. 
}
\label{fig:trigmap}
\end{figure*}

\section{Effect of Milky Way on linear polarization of 21 cm}
\label{sec:results}
In this section, we investigate how the linear polarization of 21 cm
signal described in Sec.~\ref{sec:emodepol} gets modified as it
propagates through the Galaxy. In our calculation, we assume that, during
the EoR, only E-mode polarization was created which then gets transformed
into E-mode and B-mode due to the Galactic Faraday rotation according to
the equations described in Sec.~\ref{sec:frbasics}.

Faraday rotation in Milky Way is measured by observing
extragalactic polarized radio sources.
Oppermann et al. have released publicly available rotation measure (RM) map of Milky
Way~\cite{Oppermann:2011td}. They have reconstructed the RM maps
based on the RM catalog of the NRAO VLA Sky Survey~\cite{Condon:1998iy} with several other catalogs.
In this paper, we adopt their public data as the Galactic RM map. 
The RM and its uncertainty are strongest along the Galactic
disk. Therefore, the Galactic disk is
the source of highest amount of foreground.
We choose to remove this area for our calculations, by adopting Planck's
mask 30~GHz~\cite{planck:maps}. 

In Fig.~\ref{fig:anglemap}, we present the angle of rotation of
linear polarization of 21~cm radiation based on the data of
Ref.~\cite{Oppermann:2011td} as it passes through the Galaxy. Due to the rest-frame low
frequency of these photons, the rotation angle is high. The top panel
of Fig.~\ref{fig:anglemap} shows the rotation angle in radian. The bottom panel indicates fractional uncertainty in the rotation angle. From
Sec.~\ref{sec:frbasics}, Faraday
rotation angle is proportional to square of the observed wavelength.
Therefore rotation angle increases as $(1+z)^2$ with redshift.
For the particular case of Fig.~\ref{fig:anglemap}, we assume that 21~cm
radiation was emitted at $z=$12.
 From the magnitude of the rotation angle it is
evident that there are regimes with angle of rotation being
significantly larger than 2$\pi$ radian. This creates a degeneracy in the rotation
angle or the RM along the line of sight. In order to reconstruct the
initial polarization, one should therefore be careful since along 
many lines of
sight the linear polarization may have been rotated by an angle
larger than 2$\pi$.

In Fig.~\ref{fig:trigmap}, we present cosine and sine of twice the
angle of rotation due to the Galactic magnetic field. 
We chose to plot twice the angle of rotation following the equations in
Sec.~\ref{sec:frbasics} which involves a trigonometric factor of
$2\alpha$ where $\alpha$ is the Faraday rotation angle. The top panel of Fig.~\ref{fig:trigmap} indicates full strength of Faraday rotation spanning the entire
range between $-1$ and 1. The mid panel of Fig.~\ref{fig:trigmap}
shows cosine and sine of twice the angle of Faraday rotation when
only 10\% RM is used. We parameterize the chosen RM strength by a factor $f=0.1$. The bottom
panel displays the rotation angle when only 1\% RM was used or $f=0.01$.  
It is useful to
investigate maps of $\sin(2f\alpha)$ or $\cos(2f\alpha)$ for the
following reason. If the Galactic RM is known over all sky or over
certain patches patches of the sky with sufficient accuracy (with less
than 1~\% error), then the information about the initial E-mode
polarization of 21~cm photons can be recovered. The polarization
information is completely lost due to the large rotation suffered by
21~cm photons emitted at $z \sim z_{r}$ as it passes through our
Galaxy. Therefore, if Galactic RM is known with sufficient accuracy
along several line of sights and RM is comparatively low along those
lines of sights, the initial linear polarization of
21~cm photons could be recovered along those lines. This conclusion is based on the
assumption that there is no other foreground.

We construct $C_{\ell}^{\sin(2\alpha)}$ and
$C_{\ell}^{\cos(\alpha)}$ from the
observed FR map. The Galactic FR map has a strong latitude dependence.
We simply define the rotation angular spectrum as
\begin{equation}
C^{\alpha \alpha}_\ell
\equiv (2\ell+1)^{-1} \sum_m \alpha^{\rm map*}_{\ell m}\alpha^{\rm
map}_{\ell m}. 
\end{equation}
Similarly, we obtain 
\begin{eqnarray}
C^{\sin(2\alpha) }_\ell &=& (2 \ell+1)^{-1} \sum_m (\sin 2\alpha)^{\rm map*}_{\ell m}(\sin 
2\alpha)^{\rm map}_{\ell m},
\nonumber \\
C^{\cos(2\alpha) }_\ell &=& (2 \ell+1)^{-1} \sum_m (\cos 2\alpha)^{\rm map*}_{\ell m}(\cos 
2\alpha)^{\rm map}_{\ell m}. 
\end{eqnarray}

\begin{figure}[tbp]
\includegraphics[height=0.42\textwidth]{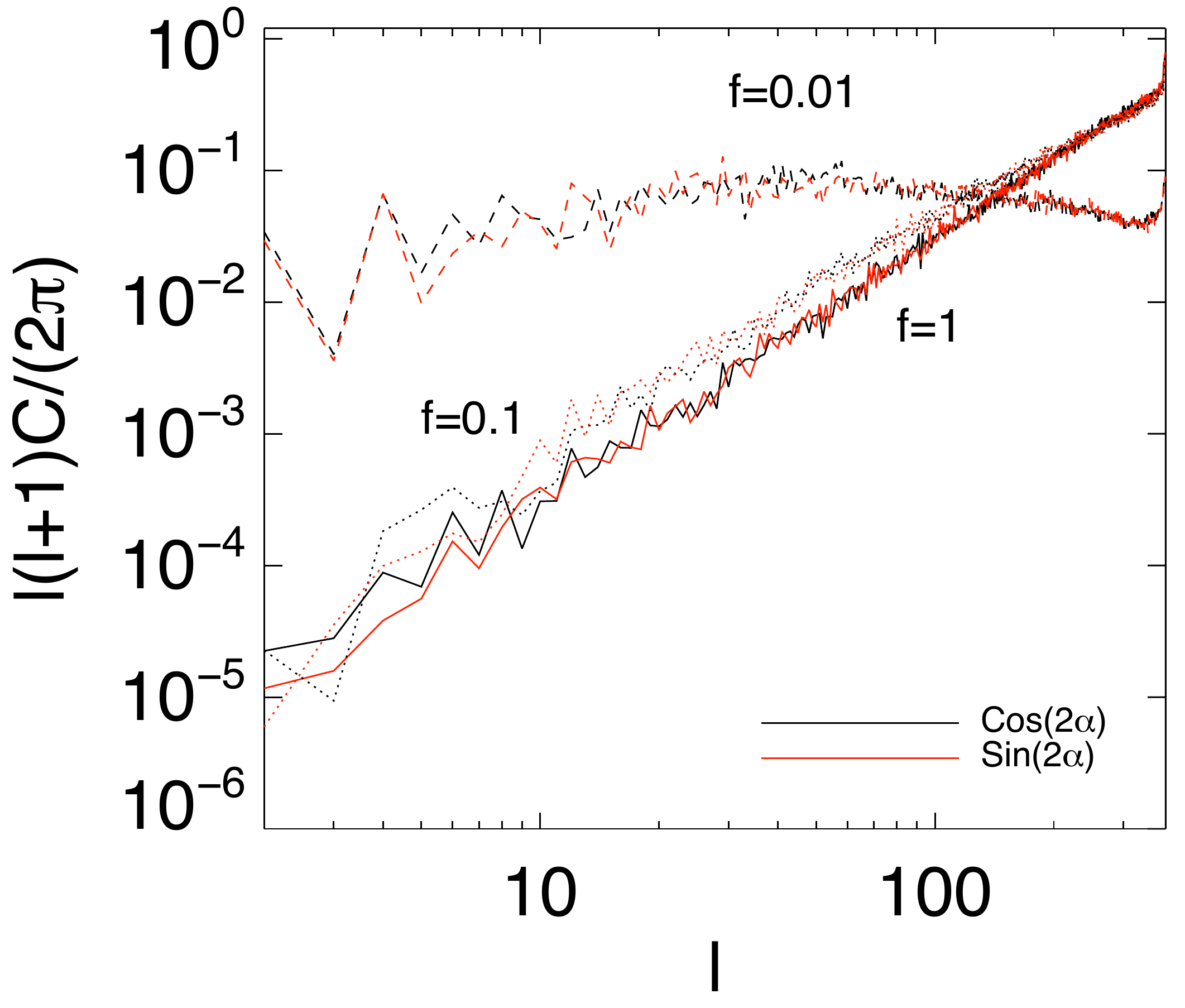}
\caption{
Power spectra of the sine and cosine of the rotation angle corresponding
to Faraday rotation of 21~cm photons emitted and polarized at $z=$12
with different levels of RM strength.}
\label{fig:rmpower}
\end{figure}

We construct $C_{\ell}^{\sin(2\alpha)}$ and
$C_{\ell}^{\cos(\alpha)}$ from the
observed FR map. In Fig.~\ref{fig:rmpower}, we present the power spectra $\ell(\ell+1)C^{\rm
RM}_{\ell}/2\pi$ of $\sin(2f\alpha)$ and $\cos(2f\alpha)$ where $\alpha$ is
the rotation in linear polarization of rest-frame 21~cm photons
emitted at redshift $z=12$. The solid,
dotted and dashed lines respectively indicate $f=$1, 0.1, 0.01. Each of
these power spectra are calculated upon removing the Galactic disk by
Planck mask cut at $70$ GHz which corresponds to a $f_{\rm sky}=0.7$. 
We notice from Fig.~\ref{fig:rmpower} that corresponding to $f=0.01$,
both sine and cosine of $2f\alpha$ reach almost a scale-invariant
shapes. This scale invariance breaks down with $\ell > 200$ corresponding to
around 0.6 degree. For higher values of $f$, we find that the power due to
$\sin(2f\alpha)$ or $\cos(2f\alpha)$ to be $\sim$ $\ell^{2}$ implying a
white-noise like power in $C_{\ell}$. We find that, upon reducing $f$ upto
0.001, the power in $\sin(2f\alpha)$ starts to resemble RM power spectra
in~\cite{Decmb2013} still remaining largely scale-invariant. 
At this point, power in $\cos(2f\alpha)$ becomes
insignificant. So far we have explored quantitative nature of Faraday
rotation angle due to our Galaxy on 21~cm polarization.  Next, we investigate the effect of FR in E-mode
polarization level originating from the EoR.

\begin{figure}[tbp]
\includegraphics[height=0.40\textwidth]{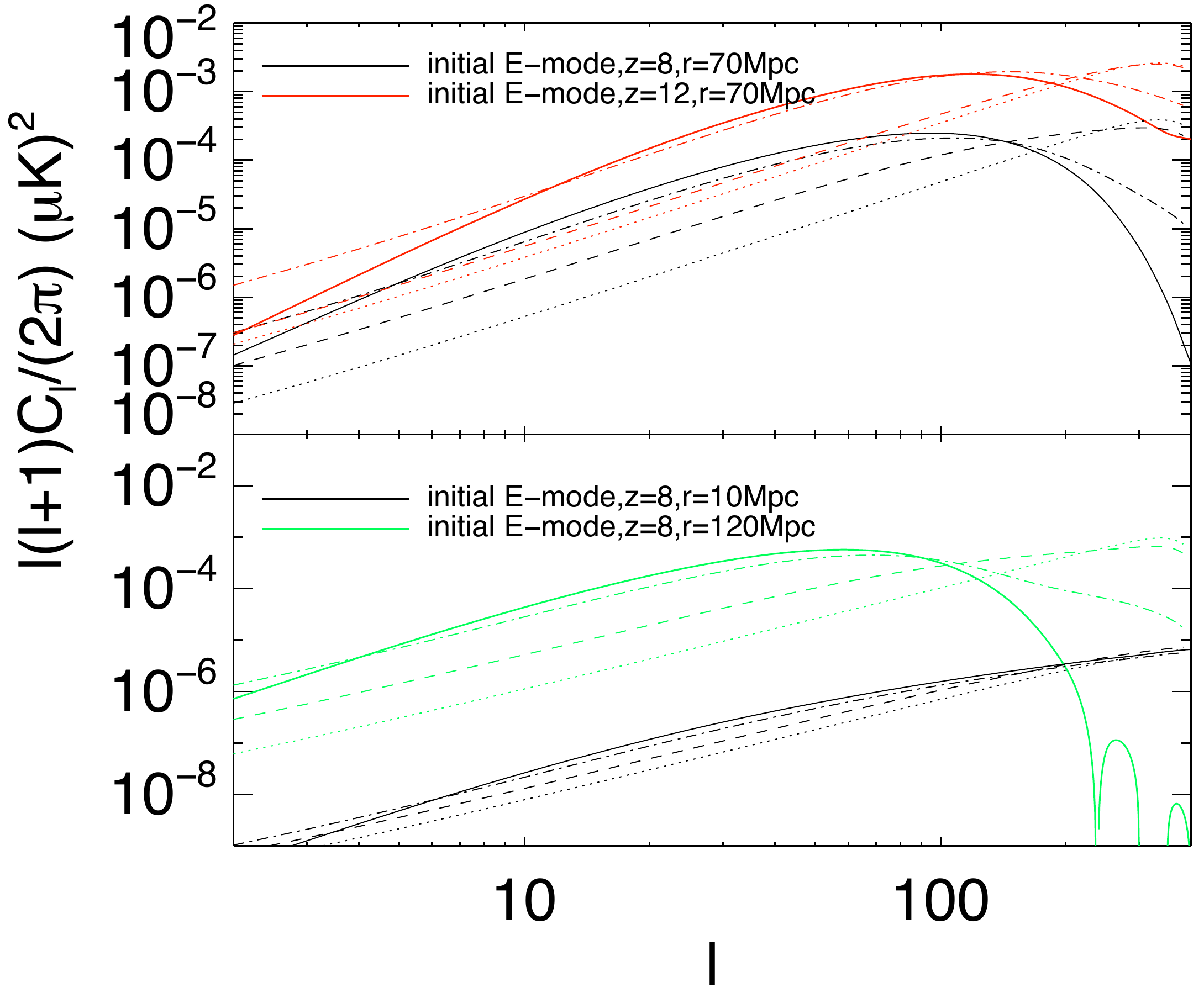}
\caption{
E-mode polarization spectra of 21~cm photons around the EoR. The solid lines
represent the initial E-mode power at the indicated
redshifts. The dotted, dashed and dot-dashed lines show E-modes
generated when RM is 100,10 and 1 percent of the actual magnitude.}
\label{fig:clee}
\end{figure}

In Fig.~\ref{fig:clee}, we present E-mode polarization of 21~cm radiation coming
from the EoR. On the top panel, the initial E-mode angular power
spectra are by solid lines. Setting that the observation redshift corresponds to the
reionization redshift, $z=z_r$, we choose two redshifts, $z=$8, 12 respectively indicated by black
and red solid lines. In the top panel, we choose a bubble radius, $r=$70 Mpc. With
dotted, dashed and dot-dashed lines we present the transformed
E-mode polarization after FR due to the Milky way with
different choices of RM strengths. Dotted lines
show when full strength of FR was applied. Dashed and dot-dashed
lines respectively indicate when only 10 and 1\% RM was used to
transform the initial E-modes according to Eq.~(\ref{eq:somae}). 
The bottom panel of Fig.~\ref{fig:clee} shows how the initial E-mode
polarization power transforms under FR when different bubble radii were
chosen. The initial polarization is
chosen to have formed at $z=$8. The black and green solid lines are chosen
respectively to represent bubble radii of 10 and 120 Mpc respectively. Similar to the
top panel, the dotted, dashed and dot-dashed lines represent 100, 10 and
1\% RM strength.

 Our conclusion from Fig.~\ref{fig:clee} is that with $f=0.01$ (with 1\%
of RM strength) one may be able to reconstruct the initial E-mode
spectra (upto $\ell \sim 100$), especially around the peak scale, in the absence of any other foreground.
The peak scale corresponds to the typical size of ionized bubbles. 
Therefore, we can access the evolution of the ionized bubble size with $f=0.01$.

\begin{figure}[tbp]
\includegraphics[height=0.40\textwidth]{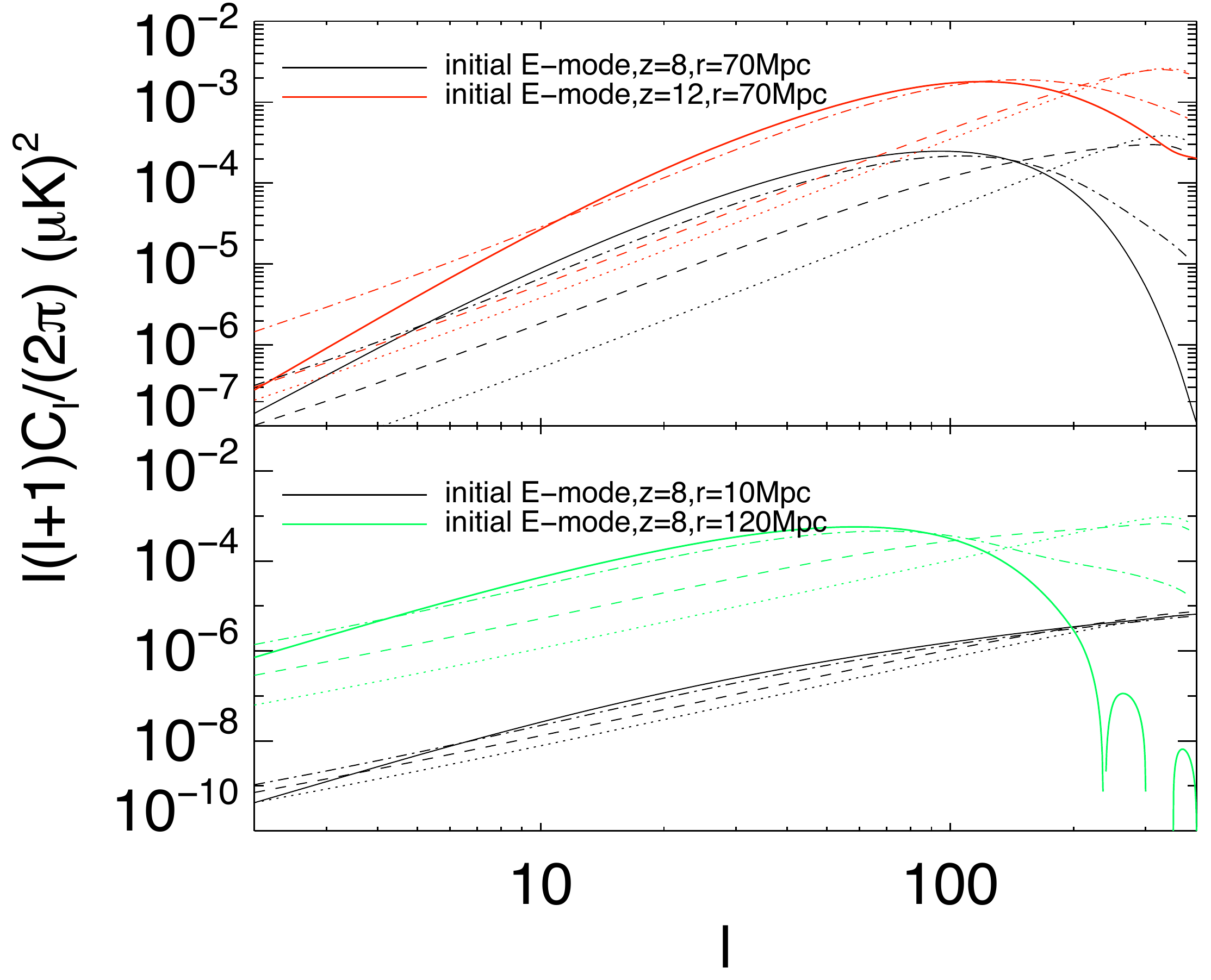}
\caption{
B-mode polarization spectra of 21~cm photons after they pass through the
Milky way. Solid lines are for the initial E-mode at the indicated
redshifts. The dotted, dashed and dot-dashed lines represent B-modes
generated when RM is 100,10 and 1 percent of the actual strength.}
\label{fig:clbb}
\end{figure}

In Fig.~\ref{fig:clbb}, we present how the E-mode polarization of 21~cm radiation coming
from the EoR transforms into B-modes due to FR from the Milky way.
On the top panel,
the initial E-mode angular power spectra for $z=8$ and $12$ are
indicated by black and red solid lines, respectively.
We choose $r=70$~Mpc for a bubble radius. 
With
dotted, dashed and dot-dashed lines,
we represent the transformed
B-mode polarization after FR due to the Milky way. Dotted lines
indicate when the full strength of RM was applied. Dashed and dot-dashed
lines respectively are for the cases where only 10 and 1\% RM was used to
transform the initial E-modes into B-modes 
according to Eq.~(\ref{eq:somab}). 
The bottom panel of Fig.~\ref{fig:clbb} shows how the initial E-mode
polarization power spectrum transforms into B-modes under FR for
different bubble radius.
The initial polarization is
chosen to have formed at $z=$8 with $z_r=8$. The black and green solid lines are chosen
respectively to present bubble radii of 10 and 120~Mpc. Similar to the
top panel the dotted, dashed and dot-dashed lines indicate 100,10 and
1\% FR strength. Similar to transformation of E-modes into
E-modes (as described in Fig.~\ref{fig:clee}), we find that B-modes
tend to replicate the initial E-modes if $f<$0.01 upto $\ell< 200$.

We investigate if there is a smooth
area on the Galactic sky where the amount of RM is small and also known
with 99\% accuracy (see Fig.~\ref{fig:trigmap}). Given the current status of FR data, we find that
there are only a few scattered lines of sights with 
such desired accuracy and
therefore construction of a smooth map is not possible among those points.

\section{Summary and Outlook}
\label{sec:summary}
The goal of this paper is to examine to what extent FR due to the Galaxy
affects the initial E-mode polarization of the 21~cm radiation and
investigate if it is  possible at all to recover the initial signal given the current
accuracy of FR data. 21~cm radiation usually unpolarized and picks up a
linear E-mode polarization during the EoR from Thomson scattering. In our
calculation we investigate how this linear polarization transform under
the Galactic FR.

We exclude any other additional contributions to RM outside of our
Galaxy. This additional contribution could come from the IGM with large
scale magnetic field of a few nG over a large path length. There could
also be contribution due to locally strong magnetic field, for
example, due to a quasar along the line of sight. In our paper we exclude such
effects and consider the effect of only the large scale magnetic field
due to the Milky way on the E-mode polarization of 21~cm radiation from
the EoR.

From Fig.~\ref{fig:trigmap}, we notice that
given the full strength of FR due to only our Galaxy, angle of rotation
of the polarization corresponding to 21~cm radiation coming from
the EoR gets rotated by a large angular range (larger than 2$\pi$). This destroys
the initial signal entirely. From the bottom panel of the same
Figure, we see that some signal may be recovered if the FR data is known
within 99\% accuracy. In Fig.~\ref{fig:anglemap}, we present the
uncertainty in the current FR map from Oppermann dataset.
We find that given this existing FR catalog based on 
the extragalactic polarized radio sources, there are only a few scattered points
where the desired accuracy to recover the initial E-mode signal is achieved.
Construction of a smooth map where the Galactic RM is known with 99\% is not
possible given the current status of observations.
Since the EoR redshift range is not precisely
determined, one may choose a lower redshift (for example $z=6$) reducing
the amount of rotation. This choice, however, reduces the initial E-mode
signal~\cite{babich21cm2005}. Therefore we conclude that given the
current status of extra-galactic  dataset mapping out the Faraday sky,
it is not possible to reconstruct the initial E-mode polarization of the
21~cm signal originating from the EoR.

However, Low Frequency Array (LOFAR)~\cite{LOFAR} and the next generation of radio survey, Square Kilometre Array
(SKA)~\cite{SKA}, is expected to improve Galactic RM data by their pulsar surveys~\cite{2011AIPC.1381..117B}.
Such improvement may allow us to construct a smooth map within 99\%
accuracy and access the polarization
signal of 21~cm radiation from the EoR.

\acknowledgments We are grateful to Niels Oppermann and
collaborators~\cite{Oppermann:2011td} for making their rotation measure
maps publicly available at \cite{RMmaps}. Some of the results in this
paper have been derived using the HEALPix \cite{healpix,Gorski:2004by}
package. We thank Levon Pogosian and Tanmay Vachaspati for an earlier
collaboration which was a motivation for this work. We also thank
Yin-Zhe Ma and Andrew Long for discussions and helpful comments. SD is supported by a
NASA Astrophysics Theory Grant NNX11AD31G and HT is supported by the DOE
at ASU.

\newpage

\end{document}